\documentclass[a4paper, amsfonts, amssymb, amsmath, reprint, showkeys, nofootinbib, twoside]{revtex4-1}
\usepackage[english]{babel}
\usepackage[utf8]{inputenc}
\usepackage[colorinlistoftodos, color=green!40, prependcaption]{todonotes}
\usepackage{amsthm}
\usepackage{ulem}
\usepackage{multirow} 
\usepackage{xcolor}
\usepackage{graphicx}
\usepackage{comment}
\usepackage{caption}
\usepackage[T1]{fontenc}
\usepackage[pdftex, pdftitle={Article}, pdfauthor={Author}]{hyperref} % For hyperlinks in the PDF

\definecolor{purple}{rgb}{0.58,0.0,0.83}

\definecolor{blue(pigment)}{rgb}{0.2, 0.2, 0.6}

\usepackage{scalerel}
\usepackage{tikz}
\usetikzlibrary{svg.path}

\definecolor{orcidlogocol}{HTML}{A6CE39}
\tikzset{
  orcidlogo/.pic={
    \fill[orcidlogocol] svg{M256,128c0,70.7-57.3,128-128,128C57.3,256,0,198.7,0,128C0,57.3,57.3,0,128,0C198.7,0,256,57.3,256,128z};
    \fill[white] svg{M86.3,186.2H70.9V79.1h15.4v48.4V186.2z}
                 svg{M108.9,79.1h41.6c39.6,0,57,28.3,57,53.6c0,27.5-21.5,53.6-56.8,53.6h-41.8V79.1z M124.3,172.4h24.5c34.9,0,42.9-26.5,42.9-39.7c0-21.5-13.7-39.7-43.7-39.7h-23.7V172.4z}
                 svg{M88.7,56.8c0,5.5-4.5,10.1-10.1,10.1c-5.6,0-10.1-4.6-10.1-10.1c0-5.6,4.5-10.1,10.1-10.1C84.2,46.7,88.7,51.3,88.7,56.8z};
  }
}

\newcommand\orcidicon[1]{\href{https://orcid.org/#1}{\mbox{\scalerel*{
\begin{tikzpicture}[yscale=-1,transform shape]
\pic{orcidlogo};
\end{tikzpicture}
}{|}}}}

\begin{document}

\title{The holographic origin of future singularities and the role of spatial curvature in cosmic expansion
%Big Rip or Little Rip? The Fate of the Universe in Curved Holographic Cosmologies with Kaniadakis Entropy
}

\author{Miguel Cruz$^1$\orcidicon{0000-0003-3826-1321}}
\email{miguelcruz02@uv.mx}

\author{Samuel Lepe$^{2}$\orcidicon{0000-0002-3464-8337}}
\email{samuel.lepe@pucv.cl}

\author{Joel Saavedra$^2$\orcidicon{0000-0002-1430-3008}}
\email{joel.saavedra@pucv.cl}

\affiliation{$^1$Facultad de F\'{\i}sica, Universidad Veracruzana 91097, Xalapa, Veracruz, M\'exico,\\
$^2$Instituto de F\'\i sica, Pontificia Universidad Cat\'olica de Valpara\'\i so, Casilla 4950, Valpara\'\i so, Chile.}

\date{\today}

\begin{abstract}
We investigate the fundamental cosmological implications of holographic dark energy using the Granda-Oliveros (GO) infrared cutoff, spatial curvature, and generalized entropies. We demonstrate that the GO cutoff establishes a geometric origin for phantom acceleration, inevitably leading to a big rip singularity without requiring exotic matter. Incorporating spatial curvature reveals that topology acts as a quantitative catalyst; positive curvature accelerates the singularity in closed universes, but cannot alter its fundamental behavior. Furthermore, we show that Kaniadakis generalized entropy modifications are structurally insufficient to prevent this finite-time divergence. To successfully soften the big rip and yield an asymptotic little rip, it is necessary (as first alternative) to integrate irreversible thermodynamical mechanisms, such as non-equilibrium particle creation. These macroscopic processes are sufficient to {\it neutralize} the geometric divergence of the GO cutoff, as we discuss in the work.
\end{abstract}

%\pacs{98.80.Cq}

\maketitle
%%%%%%%%%%%%%%%%%%%%%%%%%%%%%%%%%%
\section{Introduction} 
%%%%%%%%%%%%%%%%%%%%%%%%%%%%%%%%%%%
The $\Lambda$CDM model has proven remarkably successful in describing the large-scale structure and evolution of the universe. However, as observational precision increases, tensions have emerged that challenge its fundamental assumptions; see, for instance, \cite{val1}. One of its pillars is the flatness condition for the universe ($\Omega_k = 0$), a consequence of inflationary theory that has long served as an a priori assumption in contemporary cosmology. Although combined datasets such as Cosmic Microwave Background (CMB) radiation paired with Baryon Acoustic Oscillations (BAO) generally support flat geometry, recent analyzes of Planck 2018 power spectra alone have revealed a preference for a closed universe ($\Omega_k < 0$) at a statistical significance exceeding 99\% confidence limits \cite{planck, val2}. See also \cite{dev}, which measures the spatial curvature by exploiting departures from statistical isotropy induced by the Alcock-Paczy\'nski effect in large-scale galaxy clustering.

This curvature tension is quantitatively significant; for instance, in Ref. \cite{val2}, a constraint on $\Omega_k$ that deviates from flatness by more than 3 standard deviations was reported. This preference for positive curvature is driven by an anomalous lensing amplitude in the CMB power spectra, which is more accurately described by a closed geometry than by a flat $\Lambda$CDM model. However, this result disagrees with measurements from local cosmological probes. When Planck data is combined with BAO measurements or Type Ia Supernovae data (such as the Pantheon sample), the evidence shifts back towards a flat universe, creating a discordance that some authors argue constitutes a {\it crisis in cosmology}. 

Further studies suggest that the assumption of a flat universe might mask these discrepancies, requiring a curved background to properly evaluate the consistency of datasets \cite{cur, col}. Similarly, in \cite{cur2} it was emphasized that, although non-CMB data on their own are compatible with a flat universe, their combination with CMB data can yield conflicting constraints depending on the specific parameterizations of dark energy. Recent observational developments have further strengthened the motivation to revisit scenarios with non-vanishing curvature. Combined analyzes involving Planck, BAO, SNe, and more recently DESI DR2 have shown mild but persistent deviations from perfect flatness once extensions beyond $\Lambda$CDM are considered, together with renewed tension in the Hubble parameter $H_0$ and the clustering amplitude $S_8$. These indications suggest that spatial curvature may play a non-trivial dynamical role, especially in late-time cosmology. As emphasized in \cite{rec}, the curvature contribution has to be included to prevent inaccuracies in reconstructing a variable parameter state that characterizes a dynamical dark energy sector.

Therefore, considering the curvature parameter $k$ as a dynamical quantity instead of fixing it to zero is justified not only on theoretical grounds but also by current observations. If the universe possesses a non-trivial topology, the resulting geometric contribution to the Friedmann equations could profoundly reshape our understanding of the cosmic expansion history and its ultimate fate.

Alongside these geometric discussions lies the unresolved question concerning the true nature of Dark Energy. Whereas $\Lambda$ remains the standard explanation, recent observational developments have pointed the search toward dynamical alternatives. In particular, the first year results from the Dark Energy Spectroscopic Instrument (DESI) have provided intriguing hints that the dark energy equation of state may evolve with time ($\omega \neq -1$), challenging the constant dark energy paradigm \cite{desi}. These findings, arising from combining DESI BAO data with CMB and Supernova datasets, show a preference for dynamical behavior that naturally aligns with the predictions of time-varying dark energy models. Among the candidates proposed to replace $\Lambda$, HDE stands out as a viable alternative. Based on the holographic principle of quantum gravity \cite{qg1, qg2}, HDE postulates that the entropy-area relationship of the cosmic horizon determines the energy density of the dark sector. However, adopting the Hubble horizon as the infrared cutoff in its usual form does not yield an accelerating universe, necessitating the introduction of modified cutoff prescriptions.

The GO cutoff \cite{Granda:2008dk}, which depends on both the Hubble parameter $H$ and its time derivative $\dot{H}$, successfully generates acceleration and resolves the causality issues inherent in previous models. This specific cutoff is consistent with observational constraints and provides a rich phenomenological structure for late-time cosmology \cite{manosh}. On the other hand, a recent study suggests that the GO model may lead to the emergence of holographic dark matter, which in turn naturally accounts for the observed coincidence between baryonic and non-baryonic components \cite{holodark}.

Furthermore, the thermodynamic foundations of HDE are susceptible to modification. The standard Bekenstein-Hawking entropy, $S \propto A$, assumes standard Boltzmann-Gibbs statistics. In the context of relativistic systems and long-range interactions with non-ergodic properties, generalized entropies may offer a more accurate description. The Kaniadakis entropy (or $K$-entropy) is a one-parameter generalization arising from relativistic statistical theory \cite{kaniadakis}, and has recently been applied to cosmological horizons. These entropic corrections modify the Friedmann constraints, introducing new terms that effectively behave as dark energy and can drive the universe to future singularities \cite{kbr}. These additional contributions can significantly modify the evolution of the dark sector and, in particular, the nature of late-time instabilities. Understanding how such entropy-driven corrections interplay with spatial curvature is therefore essential for building a consistent holographic cosmology. See also Ref.~\cite{usk}, which shows that Kaniadakis cosmology successfully reproduces the behavior of a cosmological constant. This work investigates the interplay among the three components described above: non-zero spatial curvature, the GO holographic cutoff, and Kaniadakis entropy corrections. Previous research has generally investigated these effects separately; this gap provides a central motivation for the present analysis. We demonstrate that the spacetime topology acts as a catalyst for cosmic expansion, altering the onset of the phantom regime while leaving the character of the singularity unchanged. This is preserved for cases such as the big rip and the little rip. $8\pi G = k_{\mathrm{B}}=c = 1$ unit will be used in our analysis on the framework of a Friedmann–Lemaître–Robertson–Walker (FLRW) spacetime with metric given by $ds^{2} = -dt^{2}+a^{2}(t)\left[dr^{2}/(1-kr^{2})+r^{2}(d\theta^{2}+\sin^{2}\theta d\varphi^{2}) \right]$ with spatial curvature $k = 0, \pm 1$ and $a(t)$ being the scale factor, as usual.

The work is organized as follows: Section \ref{sec:cm} presents the general features of the GO cutoff and discusses the big rip singularity that arises within this model, as well as the impact of spatial curvature on cosmological evolution. In Section \ref{sec:kaniadakis}, we broaden our analysis by incorporating the effects of Kaniadakis entropy in the description of HDE and derive cosmological solutions to investigate the role of spacetime topology. Section \ref{sec:little} is dedicated to examining the conditions for the little rip scenario within our framework. Finally, in Section \ref{sec:final}, we summarize and comment on the main results of our work.

%%%%%%%%%%%%%%%%%%%%%%%%%%%%%%%%%%%%%%%%%%%%%%%%%%%%%%%%%%%
\section{Curvature parameter and future singularity}
\label{sec:cm}
%%%%%%%%%%%%%%%%%%%%%%%%%%%%%%%%%%%%%%%%%%%%%%%%%%%%%%%%%%
The well-known Granda and Oliveros model for HDE is given in terms of the Hubble scale $H$, and its first derivative as follows \cite{Granda:2008dk}\footnote{In Ref. \cite{cohen} Cohen et al., argued that in quantum field theory a short-distance cutoff is linked to a long-distance cutoff through the constraint imposed by black hole formation. Specifically, if $\rho$ denotes the quantum zero-point energy density with a short-distance cutoff, then the total energy contained in a region of size $L$ must not exceed the mass of a black hole of radius $L$. This requirement leads to $\rho  L^{3} \leq L$, which in turn implies $\rho = 3c^{2}L^{-2}$, where $3c^{2}$ is a constant introduced for convenience and $L$ is the infrared cutoff. Therefore, the GO cutoff is given by $L=\left(3\beta_{1}H^{2}+3\beta _{2}\dot{H}\right)^{-1/2}$, leading to the energy density (\ref{eq:go1}). It is worth noting that the inequality $\rho L^{3} \leq L$ can be rewritten as $\rho L^{4} \leq S_{\mathrm{BH}}$, where $S_{\mathrm{BH}}$ denotes the Bekenstein–Hawking entropy, given by $A/4$, with $A$ being the area proportional to $L^{2}$. Therefore, the entropy associated with the energy density (\ref{eq:go1}) is the usual Bekenstein–Hawking one. Any change in the entropy leads to corresponding changes in the energy density. As an illustration, we consider the Barrow entropy \cite{barrow}, for which $S_{\mathrm{B}}\propto A^{1+\Delta/2} \propto L^{2+\Delta}$. Under this assumption, using $\rho L^{4} \leq S_{\mathrm{B}}$, we find that the energy density expressed in terms of the infrared cutoff behaves as $\rho \propto L^{\Delta-2}$, where taking the limit $\Delta=0$ reproduces the standard result.}
\begin{equation}
    \rho _{de}=3\left(\beta_{1}H^{2}+\beta _{2}\dot{H}\right). \label{eq:go1}
\end{equation}
In their work \cite{Granda:2008dk}, Granda and Oliveros showed that the range $\beta_2 \,\sim\, 0.5-0.7$ is consistent with observational data and yields a physically plausible cosmic history, namely a transition from decelerated to accelerated expansion. Moreover, the value of $\beta_1$ can always be expressed in terms of $\beta_2$ using the normalization condition derived from the Friedmann constraint, then the parameters $\beta$ satisfy the condition $0<\beta _{1,2}<1$. From now on derivatives with respect to cosmic time will be represented by dots. This proposal extends the standard HDE model, substituting the future event horizon with the Ricci scalar to circumvent causality issues; see Refs. \cite{riccia, riccib}. It should be emphasized that the GO cutoff introduces a local geometrically motivated dependence on $(H,\dot H)$, making the resulting dynamics particularly sensitive to curvature compared to traditional HDE prescriptions. Therefore, the inclusion of spatial curvature is expected to produce qualitative changes in the resulting phase-space trajectories, especially near potential future singularities. As already known, for a non-flat FLRW universe the Ricci scalar is simply $R=6\left(2H^{2}+\dot{H}+k/a^{2}\right)$. Accordingly, we broaden the infrared GO cutoff defined above in the following way   
\begin{equation}
    \rho_{de} = 3\left(\beta_{1}H^{2}+\beta_{2}\dot{H}+\beta_{3}\frac{k}{a^{2}}\right),\label{eq:go}
\end{equation}
where the infrared cutoff in our formulation is given as $L=\left(3\beta_{1}H^{2}+3\beta_{2}\dot{H}+3\beta_{3}\frac{k}{a^{2}}\right)^{-1/2}$. Based on the above discussion we can once more impose the following condition on the model’s free parameters $0<\beta _{1,2,3}<1$. In order to study the effects of the curvature parameter on cosmological evolution, the Friedmann constraint will be given as follows
\begin{equation}
3H^{2}=3\left( \beta _{1}H^{2}+\beta _{2}\dot{H}\right) -\frac{3\mathbf{k}}{a^{2}}, \label{eq:withk}
\end{equation}
where the curvature parameter is denoted by $k$, as usual and we also have defined $\mathbf{k}\equiv (1-\beta_{3})k$ leading to $0<|\mathbf{k}|<1$. As can be seen, the additional contribution arising from the extended model (\ref{eq:go}) is straightforwardly incorporated via the curvature term in the Friedmann constraint. Thus, in the case of zero curvature, $k = 0$, this additional term disappears. Note that the curvature term behaves as an effective energy density scaling as $a^{-2}$, intermediate between radiation and matter, and therefore may dominate over the holographic term during certain stages of expansion, even if $|\mathbf{k}|$ is small. This sensitivity is particularly relevant in the GO scenario, where local geometric quantities drive the dynamics.

For comparative purposes, we first analyze the cosmological evolution for the flat case given by $k=0$, the deceleration parameter, $q\equiv -1 -\dot{H}/H^{2}$, is simply
\begin{equation}
q=-1-\left( \frac{1-\beta _{1}}{\beta _{2}}\right), \label{eq:alw}
\end{equation}
which is constant, and due to the definition of the constants $\beta_{1,2}$, we will always have $q<-1$, which represents a phantom evolution. This result highlights an important feature of the GO model: the emergence of phantom-like acceleration is purely geometric and does not require the introduction of exotic matter fields with $\omega<-1$.

From the Friedmann constraint (\ref{eq:withk}), we can establish the following differential equation for the Hubble parameter
\begin{equation}
\dot{H}-\frac{\left( 1-\beta _{1}\right)}{\beta_{2}}H^{2}=0, \label{eq:diff1}
\end{equation}
whose solution reads
\begin{equation}
H(t) =\left( \frac{\beta _{2}}{1-\beta _{1}}\right) \frac{1}{%
t_{s}-t}, \ \ \mbox{with}  \ \ t_{s}=t_{0}+\frac{1}{H_{0}}\left( \frac{\beta _{2}}{%
1-\beta _{1}}\right),
\end{equation}
where $H_{0}\equiv H(t=t_{0})$ denotes the current value of the Hubble constant and $t_{0}$ represents the current cosmic time. From now on, the subscript zero will stand for the evaluation of cosmological quantities at the present time. The explicit dependence $H\propto (t_s - t)^{-1}$ is characteristic of Type I singularities (big rip), for which the Hubble rate, its derivative, the energy density, and the scale factor diverge at a finite future time \cite{phan}. Significantly, the coefficient in front of $(t_s - t)^{-1}$ depends only on the ratio $(\beta_2/(1-\beta_1))$, which reinforces that the singular behavior is intrinsic to the GO structure.
%Notice that according to our solution we have $t_{s} > t_{0}$, therefore, the model exhibits a future singularity at time $t_{s}$.  
Using the definition of the Hubble parameter $H(t) \equiv \dot{a}(t)/a(t)$, one gets the scale factor as a function of cosmic time
\begin{equation}
a\left( t\right) =a_{0} \left[(\eta-1)H_{0}(t_{s}-t)\right]^{-1/\left( \eta
-1\right) },\label{eq:scale}
\end{equation}
where we have defined the constant $\eta \equiv 1+(1-\beta_{1})/\beta_{2}> 1$. The parameter $\eta$ can be interpreted as an effective measure of the {\it stiffness} of the holographic model. Larger values of $\eta$ correspond to earlier and more abrupt singularities, indicating that the ratio $(1-\beta_1)/\beta_2$ controls the severity of the phantom phase. Using the previous results, a straightforward calculation allows us to write the energy density (\ref{eq:go}) as 
\begin{equation}
\rho \left( t\right) =\frac{3}{(\eta-1)^{2}(t_{s}-t)^{2}}.
\end{equation}
It is important to note that the divergence arises even though the energy density is generated holographically through geometric quantities; thus, the big rip in the GO model is not an artifact of exotic matter, but a genuine prediction of the underlying holographic prescription. As can be observed, 
%$H\left( t\rightarrow t_{s}\right) \rightarrow \infty $, $\dot{H}\left( t\rightarrow t_{s}\right) \rightarrow \infty $, $a\left( t\rightarrow t_{s}\right) \rightarrow \infty $ and $\rho \left( t\rightarrow t_{s}\right) \rightarrow \infty$, which correspond to the defining features of a big rip singularity, as analyzed in Ref. \cite{phan}. 
these results are in agreement with the value obtained for the deceleration parameter given in Eq. \ref{eq:alw}. In a spatially flat universe where the dark energy component dominates the late-time dynamics and is governed by the HDE model (\ref{eq:go}), the cosmic evolution culminates in a big rip singularity. 

A key feature of this scenario (as well as of the scenarios analyzed in this work) is that the singularity appears to arise naturally from the geometric structure of the holographic model itself, without requiring the presence of {\it phantom} matter with $\omega < -1$, where $\omega$ denotes the parameter state of the cosmic fluid. This observation will be relevant when compared later with the Kaniadakis-corrected model, for which the singularity is softened into a little rip. The contrast between these two types of divergence is one of the central results of this work.

\subsection{$k\neq 0$}

If we now consider  $k\neq 0$, according to Eq. (\ref{eq:withk}), in analogy to the previous procedure, we can write the following expression for the first derivative of the Hubble parameter
\begin{equation}
\dot{H}=\frac{1}{\beta _{2}}\left[ \left( 1-\beta _{1}\right) H^{2}+\frac{\mathbf{k}}{a^{2}}\right], \label{eq:diff} 
\end{equation}
which leads to the following expression for the deceleration parameter
\begin{equation}
q_{k}\left( t\right) =-1-\frac{1}{\beta _{2}}\left( 1-\beta _{1}+\frac{\mathbf{k}}{a^{2}H^{2}}\right).
\end{equation}
The additional curvature-dependent term introduces a time dependence in $q_k(t)$, in sharp contrast with the flat case, see equation (\ref{eq:alw}). This sensitivity to curvature arises because the extended GO model ties the entire dark energy density to local geometric quantities including spatial curvature, which means that even small deviations from flatness can lead to detectable phenomenological differences.
%It is worth noting that once curvature is taken into account, the deceleration parameter, unlike in the case of zero curvature (see equation (\ref{eq:alw}), depends explicitly on cosmic time. 
Since the constant parameters $\beta_{1,2,3}$ obey $0<\beta_{1,2,3}<1$, the above result shows that, in a closed universe with $k=1$, the phantom behavior persists regardless of the specific values of $\beta_{1,2,3}$. Consequently, the condition $q_{1}(t)<-1$ is always satisfied for a closed expanding universe. On the other hand, for an open universe given by $k=-1$, we write  
\begin{equation}
q_{-1}\left( t\right) =-1-\frac{1}{\beta _{2}}\left[ 1-\beta _{1}-\frac{(1-\beta_{3})}{a^{2}H^{2}}\right],\label{eq:open}
\end{equation}
The threshold condition $aH = [(1-\beta_{3})/(1-\beta_1)]^{1/2}$ determines the transition between the phantom regime, a cosmological constant-like behavior, and quintessence-like dynamics. This reflects the competition between geometric expansion and curvature: in open universes, negative curvature counteracts some of the phantom behavior, delaying its onset.
%This leads to a cosmological evolution that differs from the closed universe scenario. When $aH > (1-\beta_{1})^{-1/2}$, the model exhibits phantom behavior; when $aH = (1-\beta_{1})^{-1/2}$, the HDE model mimics the cosmological constant; and for $aH < (1-\beta_{1})^{-1/2}$ the dynamics corresponds to a quintessence-type evolution. 
As the universe evolves towards $t_{s}$, it passes from one phase to another. These conditions reveal a strong relationship between the scale factor, Hubble parameter, and the constants $\beta_{1,3}$. Additionally, the phantom condition $a(t_{s})H(t_{s})\rightarrow \infty$ in (\ref{eq:open}) recovers the expression (\ref{eq:alw}), which corresponds to a phantom evolution. In summary, while a closed universe results in an evolution similar to that of a flat universe, an open universe differs in certain respects. This phase structure illustrates that spatial curvature does not remove the future singularity but instead modulates the trajectory leading to it. This will reappear in the next section, where entropy corrections (rather than curvature) will be shown to fundamentally modify the nature of the singularity.

We now turn to the cosmological evolution of a universe with non-zero spatial curvature. From Eq. (\ref{eq:diff}), we can write the following differential equation for the Hubble parameter 
\begin{equation}
\dot{H}+(1-\eta)H^{2}=\frac{\mathbf{k}}{\beta _{2}a^{2}(t)},
\end{equation}
which in turn results in
\begin{equation}
\frac{dH^{2}}{dz}+\frac{2}{1+z}\left( \eta-1\right)
H^{2}=-\frac{2\mathbf{k}/\beta _{2}}{a_{0}^{2}}\left( 1+z\right),\label{eq:hubb}
\end{equation}
expressing the dynamics in terms of redshifts casts the problem in a form well suited for comparison with observations (e.g., BAO, cosmic chronometers). In addition, the structure of Eq.~(\ref{eq:hubb}) explicitly shows how curvature contributes as an inhomogeneous driving term. As usual, the scale factor can be related to the cosmological redshift by means of transformation $1+z=a\left( t_{0}\right) /a\left(t\right) $, and we also considered $\dot{H}=-(1/2)\left( 1+z\right) dH^{2}/dz$. The solution from (\ref{eq:diff}) that describes an expanding universe can be written as
\begin{equation}
H(z) =\frac{H_{0}}{\left( 1+z\right) ^{\eta -1}}\sqrt{1+\theta
_{k}\left( 1-\left( 1+z\right) ^{2\eta }\right)},
\end{equation}
with $\theta_{k} \equiv \mathbf{k}(\eta \beta_{2}\left( a_{0}H_{0}\right)^{2})^{-1}=(1-\beta_{3})k(\eta \beta_{2}\left( a_{0}H_{0}\right)^{2})^{-1}$. The parameter $\theta_k$ acts as a dimensionless measure of the contribution of the curvature relative to the holographic energy density. The sign and magnitude of $\theta_k$ determine whether the curvature accelerates or delays the divergence of $H$. If we define the variable $x\equiv a/a_{0}$, then our solution takes the form of
\begin{equation}
    H(x)=H_{0}x^{\eta-1}\sqrt{1+\theta_{k}\left(1-x^{-2\eta} \right)},
\end{equation}
yielding,
\begin{equation}
    H_{0}\int dt=\int\frac{dx}{\sqrt{(1+\theta_{k})x^{2\eta}-\theta_{k}}}, \label{eq:int}
\end{equation}
The denominator in the integrand identifies the region where the evolution becomes singular. This geometric viewpoint establishes a direct link between the curvature-induced modification of the holographic term and the emergence of the future singularity. For $\theta_{k}>0$ (see the appendix (\ref{sec:app1}) for details), the integral has the solution
\begin{widetext}
\begin{equation}
\frac{1}{\sqrt{1+\theta_k}} \frac{x^{1-\eta}}{1-\eta} \, {}_2F_1\left(-\frac{1}{2}, \frac{\eta-1}{2\eta}, \frac{3\eta-1}{2\eta}; \frac{\theta_k}{1+\theta_k}x^{-2\eta}\right)
= H_{0}t + C,\label{eq:hyper}
\end{equation}
\end{widetext}
being $C$ an integration constant and ${}_2F_1(a,b,c;x)$ the Gauss hypergeometric function. Due to the properties of hypergeometric function, a singularity may arise when its argument takes the value one; in our case this is given by $(1+\theta_{k})x^{2\eta}-\theta_{k}=0$, as can be seen, this expression corresponds to the radicand given in the denominator of the integral (\ref{eq:int}); therefore, the condition 
\begin{equation}
    a=a_{0}\left(\frac{\theta_{k}}{1+\theta_{k}}\right)^{1/(2\eta)},
\end{equation}
leads to a singularity; note that the case $\theta_{k}<0$ must be excluded due to the exponent of the aforementioned condition. As can be seen, the value of the scale factor at which the divergence appears is determined by the constant $\theta_{k}$, which depends on the curvature parameter.  
Thus, curvature does not merely shift the singularity time, but determines the earliest admissible value of the scale factor for which the cosmological evolution remains regular. This is a direct consequence of the geometric nature of HDE within the GO prescription. Therefore, solution (\ref{eq:hyper}) exhibits a singular behavior and constrains the admissible values of the scale factor through the curvature parameter, thus ruling out the open-universe case. If we focus on the asymptotic behavior, $a\rightarrow \infty$, the integral (\ref{eq:int}) takes the form
\begin{equation}
    H_{0}\int dt \simeq \frac{1}{\sqrt{1+\theta_{k}}}\int\frac{dx}{x^{\eta}}, \label{eq:int2}
\end{equation}
from which we obtain
\begin{equation}
    a(t)=a_{0}\left[(\eta-1)\sqrt{1+\theta_{k}}H_{0}(t_{s}-t) \right]^{-1/(\eta-1)},
\end{equation}
where we have defined
\begin{equation}
    t_{s}=t_{0}+\frac{1}{(\eta-1)\sqrt{1+\theta_{k}}H_{0}}=t_{0}+\frac{1}{H_{0}\sqrt{1+\theta_{k}}}\left(\frac{\beta_{2}}{1-\beta_{1}}\right), \label{eq:sing2}
\end{equation}
this solution exhibits a future singularity and reproduces the scale factor given in (\ref{eq:scale}) when $\theta_{k}=0$. Equation (\ref{eq:sing2}) makes explicit that curvature accelerates the occurrence of the big rip: positive $\theta_k$ leads to a smaller remaining cosmic time before the singularity. This curvature-driven acceleration will later be contrasted with the Kaniadakis-modified framework, in which the singularity is mitigated rather than moved forward.

Therefore, for $k, \beta_{3} \neq 0$, the asymptotic behavior of the HDE model corresponds to a phantom-type evolution. It is important to emphasize that according to (\ref{eq:sing2}), the spatial curvature affects the moment at which the singularity occurs, since $\theta_{k}>0$, the singularity takes place earlier than in the flat case. 
This means that the presence of $\theta_{k}$ speeds up the expansion, bringing the singularity closer in a shorter time. The topology of spacetime acts as a catalyst of cosmic expansion within this holographic description of the dark energy. This observation naturally motivates the search for mechanisms capable of softening or removing the big rip within holographic scenarios. As we show in the next section, Kaniadakis entropy corrections provide precisely such a mechanism by modifying the infrared structure of the holographic energy density.

To end this section, we comment on the standard scenario for a phantom fluid with constant parameter state $\omega <-1$ plus curvature.  The scale factor is also solved from the Friedmann constraint $3H^{2} = \rho - k/a^{2}$ together with $\rho=\rho_{0}a^{-3(1-|\omega|)}$, then
\begin{equation}
    \int\frac{dx}{\sqrt{x^{\eta}-\theta_{k}}} = \sqrt{\frac{\rho_{0}}{3}}\int dt, \label{eq:stand}
\end{equation}
where $\eta \equiv 3|\omega|-1$ and $\theta_{k}\equiv 3k/(a^{2}_{0}\rho_{0})$. In this case, we have the following expression
\begin{equation}
\frac{2x^{(2-\eta)/2}}{2-\eta} \, {}_2F_1\left(-\frac{1}{2}, \frac{\eta-2}{2\eta}, \frac{3\eta-2}{2\eta}; \theta_{k}x^{-\eta}\right)
= \sqrt{\frac{\rho_{0}}{3}}t + C,\label{eq:hyper2}
\end{equation}
which closely resembles the solution given (\ref{eq:hyper}). In this situation, the singularity arises when
\begin{equation}
    a=a_{0}\theta^{1/\eta}_{k}.
\end{equation}
This comparison shows that the influence of curvature on the future singularity is qualitatively similar in both the GO holographic model and the standard phantom-fluid scenario. However, in the GO case, the phantom behavior emerges geometrically rather than through an exotic fluid. This key difference underscores the importance of examining geometric modifications, such as Kaniadakis entropy corrections, to understand whether the big rip is a robust prediction or an artifact of the underlying entropy-area relation. Notice that in both scenarios the parameter $\eta$ encodes the properties of dark energy, while in the GO model this parameter depends on the constants $\beta_{1,2}$, in the standard phantom fluid case $\eta$ it is determined solely by its parameter state. Therefore, with respect to the future singularity, the spatial curvature plays a role similar to that in the holographic scenario since, in the limit $a\rightarrow \infty$ the integral given on the l.h.s of Eq. (\ref{eq:stand}) leads to a solution similar to the one obtained from (\ref{eq:int2}).  

%%%%%%%%%%%%%%%%%%%%%%%%%%%%%%%%%%%%%%%%%%%%%%%%%%%%%%%%%%%%%%%
\section{Extended Kaniadakis holographic dark energy}
\label{sec:kaniadakis}
%%%%%%%%%%%%%%%%%%%%%%%%%%%%%%%%%%%%%%%%%%%%%%%%%%%%%%%%%%%%%%%

The motivation for introducing Kaniadakis entropy into the HDE framework emerges from the central role played by the entropy–area relation in constructing HDE models. Any modification of the entropy, especially one motivated by relativistic, non-extensive statistical mechanics, leads to a direct generalization of the infrared behavior of the holographic bound. As a result, Kaniadakis entropy offers a coherent framework to examine whether the future singularity found in the GO setup is a genuine, robust prediction or simply a consequence of assuming the conventional Bekenstein–Hawking entropy. Additionally, the changes induced by the Kaniadakis parameter in the Friedmann equations are equivalent to a cosmological constant evolution in some regime \cite{usk}. Now we study the role of the Kaniadakis parameter in the context of HDE plus spatial curvature. The Kaniadakis entropy has the form \cite{abreu}
\begin{equation}
    S_{\mathrm{K}}=-\sum^{W}_{i=1}\frac{P^{1+K}_{i}-P^{1-K}_{i}}{2K}, \label{eq:def}
\end{equation}
being $K$ the Kaniadakis parameter, which quantifies the deviations from standard statistical mechanics; $P_{i}$ denotes the probability that the system is found in a given microstate, while $W$ represents the total number of possible configurations. In the framework of black hole physics, we take these probabilities to be uniform, so that $P_{i} = 1/W$ \cite{pavon}, using the fact that Boltzmann-Gibbs entropy is $S=\ln (W)$, it yields $W=\exp (S_{\mathrm{BH}})$, therefore we obtain from Eq. (\ref{eq:def})
\begin{equation}
    S_{\mathrm{K}}=\frac{1}{K}\sinh (KS_{\mathrm{BH}}), \label{eq:kaniadakis}
\end{equation}
where $K$ is the Kaniadakis parameter restricted to the interval, $0 < K < 1$, and $S_{\mathrm{K}\rightarrow 0}=S_{\mathrm{BH}}$, in such a limit, the Bekenstein-Hawking (BH) entropy is recovered. The hyperbolic structure of $S_K$ implies that powers of the horizon area control deviations from the standard entropy. Hence, Kaniadakis entropy naturally introduces infrared correction scaling as $L^2$, which in turn modify the gravitational dynamics at late times. As already known, $S_{\mathrm{BH}}=A/4$, being $A$ the area of the apparent horizon, $A=4\pi R^{2}_{A}$ \cite{bekenstein, haw}. If the generalized entropy (\ref{eq:kaniadakis}) is required to remain close to the standard form, then the parameter must satisfy $K \ll 1$ to ensure this condition; then Eq. (\ref{eq:kaniadakis}) takes the following form after a series expansion
\begin{equation}
    S_{\mathrm{K}} = S_{\mathrm{BH}}+\frac{K^{2}}{6}S^{3}_{\mathrm{BH}}+\mathcal{O}(K^{4}),\label{eq:kanapprox}
\end{equation}
As shown, the zeroth-order contribution corresponds to the standard entropy, while the second term represents the leading Kaniadakis correction. Consequently, taking the limit $K \rightarrow 0$ restores the BH entropy. Mathematically, the HDE model is defined by the inequality $\rho_{\mathrm{de}}L^{4}\leq S_{\mathrm{K}}$. Hence, starting from Eq. (\ref{eq:kanapprox}), we derive the dark energy density that arises from the modified Kaniadakis entropy \cite{ksari}
\begin{equation}
    \rho_{de}=3c^{2}L^{-2}+K^{2}L^{2},
\end{equation}
where $c^{2}$ and $K^{2}$ are constant parameters. The emergence of the $K^{2}L^{2}$ term is a purely entropy-driven effect: it acts as an infrared correction that becomes important only at late times, unlike the $L^{-2}$ holographic contribution. %This interplay between ultraviolet and infrared holographic components is what ultimately softens the GO big rip into a little rip, as we will see below. 
Observe that, when $K = 0$, the preceding formula reduces to the standard HDE result $\rho_{de} = 3c^{2}L^{-2}$, which corresponds to the limit mentioned above, $S_{\mathrm{K}\rightarrow 0}=S_{\mathrm{BH}}$. To make a comparison with our earlier discussion, we therefore need to take into account the infrared cutoff that was used before, then we can write
\begin{equation}
    \rho_{de} = 3\left(\beta_{1}H^{2}+\beta_{2}\dot{H}+\beta_{3}\frac{k}{a^{2}}\right)+\frac{K^{2}}{3\left(\beta_{1}H^{2}+\beta_{2}\dot{H}+\beta_{3}\frac{k}{a^{2}}\right)}.
\end{equation}
Here, we redefine the constants for convenience. By choosing $K=0$ in this last expression, we obtain the holographic model presented in (\ref{eq:go}). Once again, we start by examining the scenario of zero spatial curvature. For a universe dominated by dark energy, the Friedmann constraint can then be expressed as
\begin{equation}
3H^{2} = 3\left(\beta_{1}H^{2}+\beta_{2}\dot{H}\right)+\frac{K^{2}}{3\left(\beta_{1}H^{2}+\beta_{2}\dot{H}\right)}, \label{eq:friedk}
\end{equation}
leading to the following differential equation for the Hubble parameter
\begin{equation}
    \dot{H} = \frac{1}{\beta_2} \left[ \left(\frac{1}{2} - \beta_1\right) H^2 \pm \frac{\sqrt{9H^4 - 4K^2}}{6} \right], \label{eq:hu}
\end{equation}
observe that for $K=0$, the previous result matches the differential equation given in (\ref{eq:diff1}) only if the positive branch of the solution is selected. Then the second term given in the square brackets originates a different cosmic evolution from the GO scenario analyzed previously. In this case a de Sitter phase is allowed, the condition $\dot{H}=0$ leads to 
\begin{equation}
    H_{\mathrm{dS}} = \left[\frac{K^{2}}{9\beta_{1}(1-\beta_{1})}\right]^{1/4},
\end{equation}
which is a constant value, as expected. However, this de Sitter phase is unstable, if a perturbation of the form $H = H_{dS} + \delta H$ is inserted into (\ref{eq:hu}) gives the linearized equation
\begin{equation}
\dot{\delta H} = \left(\frac{4\beta_{1}(1-\beta_{1})}{\beta_{2}(2\beta_{1}-1)}\right)H_{dS}\,\delta H + \mathcal{O}(\delta H^{2}),
\end{equation}
indicating exponential grow for any perturbation since the term in the parenthesis is strictly positive\footnote{Considering a de Sitter evolution, from Eq. (\ref{eq:hu}) it is direct to establish the condition $(3-6\beta_{1})H^{2}_{\mathrm{dS}}=-\sqrt{9H^{4}_{\mathrm{dS}}-4K^{2}}$ implying $(3-6\beta_{1})\leq 0$ in order to have real solutions. This condition imposes $\beta_{1}\geq 1/2$. Given our definition of the parameters $\beta_i$, the value of $\beta_1$ must lie within the tighter interval $1/2 \leq \beta_1 < 1$ in order for the universe to successfully evolve into a de Sitter phase.}. Thus, any exponentially growing fluctuation $\delta H$ causes the Hubble parameter to increase and forcing the universe out of its phase of constant expansion. If we focus on the asymptotic behavior $H\rightarrow \infty$ we can perform an expansion on the Kaniadakis radical term $\sqrt{9H^{4}-4K^{2}}$, then equation (\ref{eq:hu}) takes the form
\begin{equation}
    \dot{H} \simeq \frac{1-\beta_{1}}{\beta_{2}}H^{2}-\frac{K^{2}}{9\beta_{2}H^{2}}, 
\end{equation}
and the Kaniadakis term is heavily suppressed with respect to the first one, then $\dot{H}\simeq (1-\beta_{1})H^{2}/\beta_{2}$ which corresponds to Eq. (\ref{eq:diff1}) and exhibits a big rip singularity. Turning on spatial curvature leads to analogous results; the ultimate fate of the universe remains unchanged. Consequently, the Kaniadakis entropic correction does not prevent the occurrence of a geometric big rip singularity arising from the geometric structure defined by the GO model. As the universe accelerates and $H \rightarrow \infty$, the GO cutoff approaches zero, and the geometric contributions scale as $1/L^{2}$, diverging as the universe expands. This behavior cannot be easily altered by merely adding corrective terms. An infrared modification of the type considered here does not change the ultimate fate of the universe, because in this framework the singularity arises purely from geometric reasons, as discussed in the previous section.
\section{Condition for a little rip scenario with holographic dark energy}
\label{sec:little}
For a singularity to be classified as a little rip, the scale factor and Hubble parameter must diverge as well as the cosmic time. The integral
\begin{equation}
    \Delta t = \int^{\infty}_{H_{0}}\frac{dH}{\dot{H}},
\end{equation}
must diverge \cite{lilrip}. Notice that in the GO model and the Kaniadakis-corrected one, we obtain $\dot{H} \propto H^{2}$, the integration of $1/H^{2}$ yields a finite constant yielding a big rip singularity. In order to have a divergent result for the integral we need that the time derivative $\dot{H}$ must grow no faster than linearly with respect to $H$: $\dot{H} \simeq \alpha H$. For this condition the integral becomes logarithmic and diverges as $H\rightarrow \infty$. If we consider purely the GO model plus corrections, the Friedmann constraint reads
\begin{equation}
    3H^{2} = 3(\beta_{1}H^{2}+\beta_{2}\dot{H})+\rho_{\mathrm{correc}},
\end{equation}
being $\rho_{\mathrm{correc}}$ the necessary correction term to the GO model, taking the little rip condition obtained above we obtain for the correction term
\begin{equation}
    \rho_{\mathrm{correc}}(H) = 3(1-\beta_{1})H^{2}-3\alpha \beta_{2}H.\label{eq:corr}
\end{equation}
This straightforward derivation clarifies why the Kaniadakis entropy correction alone cannot avert a finite-time singularity: a dominant term proportional to $H^{2}$ is required. Such a contribution is essential to neutralize the phantom acceleration characteristic of the GO cutoff. At high energies, the necessary correction must behave as $\mathcal{O}(L^{-2})$, instead of $\mathcal{O}(L^{2})$ as in the Kaniadakis case. An energy density like the one given in (\ref{eq:corr}) may seem nonstandard or defined for convenience. However, as we will see below, this is not the case. 
\subsection{Irreversible thermodynamics}
Standard cosmological models, including those based on HDE, implicitly assume that the universe expands adiabatically. Under this assumption, the universe is treated as a closed thermodynamic system where the total number of particles in a comoving volume remains constant, and entropy is conserved ($dS = 0$).
However, in a rapidly accelerating universe, this assumption breaks down \cite{maartens}. A detailed formulation of the cosmological description within the framework of non-equilibrium thermodynamics was provided in \cite{prigogine, zim}; in this scenario, the gravitational field is assumed to transfer energy into the matter sector, producing a negative creation pressure that effectively reproduces the behavior of dark energy with no need of extra contribution coming from an exotic type of matter. When particles are continuously created out of the gravitational vacuum, the standard conservation equation must be modified to account for the change in particle number. This introduces a new macroscopic variable into the fluid which induce deviation from the equilibrium condition: the creation pressure, $p_c$. The generalized energy conservation equation becomes:
\begin{equation}
    \dot{\rho} + 3H(\rho + p + p_c) = 0,
\end{equation}
where the creation pressure is directly proportional to the particle creation rate, $\Gamma$:
\begin{equation}
    p_c = - \frac{\rho + p}{3H} \Gamma.
\end{equation}
Since the effects of matter creation appear as an additional pressure term, a notable aspect of this type of scenario is its equivalence to models where dissipative processes are incorporated into the description of the cosmic fluid in homogeneous spacetimes, this is discussed in detail in Ref. \cite{zimdahl} from the kinetic theory perspective. If particles are being created the creation pressure $p_c$ is strictly negative since $\Gamma > 0$. If the creation rate scales dynamically with the expansion rate of spacetime (e.g., $\Gamma \propto H$) which is usually assumed, see for instance \cite{lima, waga}, the resulting creation pressure can lead to an accelerated expansion. This naturally produces energy density terms that scale precisely as $\mathcal{O}(H^2)$ and $\mathcal{O}(H)$.
This macroscopic approach resolves the fundamental tension present in the Kaniadakis-modified GO framework. The Kaniadakis entropy introduces an $\mathcal{O}(L^{2})$ correction that mathematically vanishes as $H \to \infty$, failing to prevent the geometric divergence of the big rip. Conversely, the particle creation mechanism scales \textit{with} the expansion. As the Hubble parameter grows excessively large near the singularity, the creation rate $\Gamma$ and the corresponding negative creation pressure $p_c$ grow alongside it. This provides a persistent, high-energy thermodynamic counter-force that naturally neutralizes the divergence of the GO cutoff. Therefore, a geometric singularity could be softened by enriching the thermodynamics description of the universe. Furthermore, this macroscopic thermodynamic perspective naturally contextualizes the role of spatial curvature at late times. As previously established, the geometric contribution of spatial curvature to the extended GO model scales as $k/a^2$. As the universe accelerates toward a potential singularity and the scale factor diverges ($a \to \infty$), this curvature term strictly decays to zero. Consequently, while spatial topology can act as an initial catalyst to modify the expansion rate, it vanishes at the extreme asymptotic limit, failing to alter the fundamental qualitative nature of the future singularity. Because this topological influence inevitably vanishes, any mechanism intended to sustainably soften the big rip into a little rip must emerge from the energy sector itself, decoupled from inverse scale factor dependencies.

%%%%%%%%%%%%%%%%%%%%%%%%%%%%%%%%%%%%%%%%%%%%%%%%%%%%%%%%%%%
\section{Conclusions}
\label{sec:final}
%%%%%%%%%%%%%%%%%%%%%%%%%%%%%%%%%%%%%%%%%%%%%%%%%%%%%%%%%%
The analysis of cosmological evolution under the HDE model, using the GO cutoff, confirms that, in a spatially flat universe, dynamics inevitably culminate in a big rip singularity. Incorporating spatial curvature into this standard framework demonstrates that spacetime topology plays a crucial role in evolution: a closed universe invariably maintains phantom behavior with a deceleration parameter less than $-1$, while an open universe exhibits more complex dynamics, transitioning between quintessence, de Sitter, and phantom phases. It is crucial to note that positive curvature acts as a catalyst for expansion, causing the future singularity to occur earlier than in the flat case. In this sense, the GO holographic setup provides a purely geometric route to phantom behavior and big-rip singularities, without invoking exotic fluids with $\omega<-1$. Comparison with the standard phantom fluid plus curvature shows that the qualitative role of $k$ is similar in both descriptions. However, in the GO case, the origin of the instability is encoded in the holographic cutoff itself rather than in the matter sector. Critically, extending the holographic framework with Kaniadakis or Barrow entropic corrections fails to soften this geometric singularity. Because the GO cutoff shrinks at high expansion rates, the corresponding $\mathcal{O}(L^{2})$ or $\mathcal{O}(L^{\Delta-2})$ entropic modifications either decay or actively accelerate the cosmic divergence. Our findings establish that thermodynamic deformations in the entropy-area alone are structurally insufficient to counter purely geometric phantom acceleration.
Instead, a robust transition from a big rip to an asymptotic little rip requires the inclusion of macroscopic, irreversible thermodynamical processes. Mechanisms such as continuous non-equilibrium particle creation generate a negative pressure proportional to the Hubble rate. This supplies the necessary $\mathcal{O}(H^{2})$ and $\mathcal{O}(H)$ scaling required to persistently neutralize the phantom divergence at high energies. Therefore, we conclude that while spacetime geometry dictates the phantom evolution and the curvature adjusts its timing, it is the irreversible macroscopic thermodynamics of the cosmic fluid that could ultimately lead the universe toward a different fate.

\section*{Data Availability Statement}
There are no new data associated with this article.
\section*{Conflict of interest to disclose}
 No, there is no conflict of interest. 
\section*{Acknowledgments}
M.~Cruz work was partially supported by S.N.I.I. (SECIHTI-M\'exico). S.~Lepe acknowledges the FONDECYT grant N°1250969, Chile. J.~Saavedra acknowledges the FONDECYT grant N°1220065, Chile.

\appendix
\section{Curvature and scale factor}
\label{sec:app1}
Since the values of $\eta$ and $\theta_{k}$ are arbitrary, the integral
\begin{equation}
    \int\frac{dx}{\sqrt{(1+\theta_{k})x^{2\eta}-\theta_{k}}},
\end{equation}
can be written as
\begin{equation}
    \frac{1}{\sqrt{A}}\int x^{-\eta}\left(1-\frac{B}{A}x^{-2\eta}  \right)^{-1/2}dx,
\end{equation}
where we defined the following constants: $A \equiv 1+\theta_{k}$ and $B\equiv \theta_{k}$. Compared with the integral \cite{stegun}
\begin{equation}
    \int x^{\alpha}(1-cx^{\beta})^{\gamma}dx=\frac{x^{\alpha +1}}{\alpha+1}{}_2F_1\left(\gamma, \frac{\alpha+1}{\beta}, \frac{\alpha+1}{\beta}+1; cx^{\beta}\right), \label{eq:app}
\end{equation}
we obtain $\alpha=-\eta$, $\beta=-2\eta$, $\gamma=-1/2$ and $c=B/A=\theta_{k}/(1+\theta_{k})$.Observe that the integral representation of the hypergeometric function in (\ref{eq:app}) remains valid only when $\theta_{k}>0$.

\section{Thermodynamic Entropy Modifications: The Barrow case}
To understand why purely thermodynamic modifications to the entropy-area relationship are structurally insufficient to avoid the GO divergence, as an example one can analyze the impact of the Barrow entropy. In the Barrow formulation the energy density behaves as $\rho \propto L^{\Delta-2}$. In the limit where the fractal parameter $\Delta = 0$, this reproduces the standard holographic result, as mentioned before. Applying the GO infrared cutoff $L = (3\beta_1 H^2 + 3\beta_2 \dot{H})^{-1/2}$, the Barrow-modified energy density takes the form:
\begin{equation}
    \rho_B = (3\beta_1 H^2 + 3\beta_2 \dot{H})^{1 - \Delta/2},
\end{equation}
substituting this modified density into the standard spatially flat Friedmann constraint ($3H^2 = \rho_B$) yields:
\begin{equation}
    3H^2 = (3\beta_1 H^2 + 3\beta_2 \dot{H})^{1 - \Delta/2},
\end{equation}
to determine the asymptotic expansion dynamics, we algebraically solve for the time derivative of the Hubble parameter, $\dot{H}$, we obtain the exact differential equation governing the Barrow-modified GO universe:
\begin{equation}
    \dot{H} = \frac{1}{\beta_{2}}\left[3^{\Delta/(2-\Delta)}H^{4/(2-\Delta)} - \beta_1 H^2\right].
\end{equation}
To mathematically avoid a finite-time big rip and achieve an infinite timeline characteristic of a little rip, the integral of the remaining cosmic time ($\Delta t = \int dH/\dot{H}$) must diverge as $H \to \infty$. This requires the dominant power of $H$ to be $p \leq 1$. In the standard GO framework ($\Delta = 0$), the dominant exponent is $p = 4/2 = 2$, yielding $\dot{H} \propto H^2$, which reliably forces a finite-time big rip. However, the Barrow parameter is strictly bounded by physical fractal geometry such that $\Delta \in (0, 1]$ \cite{barrow}. Consequently, the denominator $(2-\Delta)$ is strictly less than $2$, meaning the dominant exponent becomes:
\begin{equation}
    p = \frac{4}{2-\Delta} > 2.
\end{equation}
Because the dominant term dictates that $\dot{H}$ scales with a power strictly greater than $2$ at high energies, the integral $\int H^{-p} dH$ converges even more rapidly than in the standard model. Rather than preventing the singularity, the Barrow entropy modification mathematically accelerates the big rip scenario. Because the GO cutoff $L$ shrinks at high energy, the Friedmann equation forces the expansion rate $\dot{H}$ to grow exponentially faster to satisfy the modified fractal entropy bound. This rigorous asymptotic limit confirms that thermodynamic entropy-area deformations (Kaniadakis or Barrow) does not alter the nature of the GO geometric divergence. This paves the way to the necessity of integrating a macroscopic energy transfer mechanism, such as non-equilibrium particle creation (or other), to generate the appropriate $\mathcal{O}(H)$ scaling required to stabilize a big rip singularity into an asymptotic little rip.


\begin{thebibliography}{4}
\bibitem{val1}
 E.~Di~Valentino et al., Astropart.\ Phys. {\bf 131}, 102605 (2021).

\bibitem{planck}
N.~Aghanim, et~al. (Planck Collaboration), Astron.\ Astrophys. {\bf 641}, A6 (2020).

\bibitem{val2}
E.~Di~Valentino, A.~Melchiorri and J.~Silk, Nat.\ Astron. {\bf 4}, 196 (2019).

\bibitem{dev}
L.~Amendola, M.~Marinucci and M.~Quartin, Phys.\ Rev.\ Lett. {\bf 134}, 101004 (2025).

\bibitem{cur}
W.~Handley, Phys.\ Rev.\ D {\bf 103}, 041301 (2021).

\bibitem{col}
A.~A.~Coley, {\it Spatial Curvature in Cosmology Revisited}, 	arXiv:1905.04588 [gr-qc].

\bibitem{cur2}
C.-G.~Park and B.~Ratra, Astrophys.\ Space\ Sci. {\bf 364}, 134 (2019).

\bibitem{rec}
C.~Clarkson, M.~Cortes and B.~A.~Bassett, JCAP {\bf 08}, 011 (2007).


\bibitem{desi}
A.~G.~Adame, et~al. (DESI Collaboration), JCAP {\bf 02}, 021 (2025).

\bibitem{qg1}
G.~'t~Hooft, {\it Dimensional reduction in quantum gravity}, arXiv:gr-qc/9310026.

\bibitem{qg2}
L.~Susskind, J.\ Math.\ Phys. {\bf 36}, 6377 (1995).

\bibitem{Granda:2008dk}
L.~N.~Granda and A.~Oliveros, Phys.\ Lett.\ B {\bf 669}, 275 (2008).

\bibitem{manosh}
M.~T.~Manoharan, {\it Note on Granda-Oliveros Holographic Dark Energy}, arXiv:2311.12409 [astro-ph.CO].

\bibitem{holodark}
O.~Trivedi and R.~J.~Scherrer, {\it Dark Matter from Holography}, 	arXiv:2511.10617 [astro-ph.CO].

\bibitem{kaniadakis}
G.~Kaniadakis, Phys.\ Rev.\ E {\bf 66}, 056125 (2002); Phys.\ Rev.\ E. {\bf 72}, 036108 (2005).

\bibitem{kbr}
N.~Drepanou, A.~Lymperis, E.~N.~Saridakis and K.~Yesmakhanova, Eur.\ Phys.\ J.\ C {\bf 82}, 449 (2022).

\bibitem{usk}
M.~Cruz, S.~Lepe and J.~Saavedra, Nucl.\ Phys.\ B {\bf 1021}, 117190 (2025).

\bibitem{cohen}
A.~G.~Cohen, D.~B.~Kaplan and A.~E.~Nelson, Phys.\ Rev.\ Lett. {\bf 82}, 4971 (1999).

\bibitem{barrow}
J.~D.~Barrow, Phys.\ Lett.\ B {\bf 808}, 135643 (2020).

\bibitem{riccia}
X.~Zhang, Phys.\ Rev.\ D {\bf 79}, 103509 (2009).

\bibitem{riccib}
C.~Gao, F.~Wu, X.~Chen and You-Gen~Shen, Phys.\ Rev.\ D {\bf 79}, 043511 (2009).

\bibitem{phan}
S.~Nojiri1, S.~D.~Odintsov and S.~Tsujikawa, Phys.\ Rev.\ D {\bf 71}, 063004 (2005).

\bibitem{abreu}
E.~M.~C.~Abreu, J.~A.~Neto, E.~M.~Barboza and R.~C.~Nunes, EPL {\bf 114}, 55001 (2016); Int.\ J.\ Mod.\ Phys.\ A {\bf 32}, 1750028 (2017); E.~M.~C.~Abreu and J.~A.~Neto, EPL {\bf 133}, 49001 (2021).

\bibitem{pavon}
D.~Pav\'on, arXiv:2001.05716 [gr-qc].

\bibitem{bekenstein}
J.~D.~Bekenstein, Lett.\ Nuovo\ Cimento {\bf 4}, 737 (1972); Phys.\ Rev.\ D {\bf 7}, 2333 (1973); Phys.\ Rev.\ D {\bf 9}, 3292 (1974). 

\bibitem{haw}
S.~W.~Hawking, Phys.\ Rev.\ Lett. {\bf 26}, 1344 (1971).

\bibitem{ksari}
N.~Drepanou, A.~Lymperis, E.~N.~Saridakis and K.~Yesmakhanova, Eur.\ Phys.\ J.\ C {\bf 82}, 449 (2022).

\bibitem{lilrip}
P.~H.~Frampton, K.~J.~Ludwick and R.~J.~Scherrer, Phys.\ Rev.\ D {\bf 84}, 063003 (2011); I.~Brevik, E.~Elizalde, S.~Nojiri and S.~D.~Odintsov, Phys.\ Rev.\ D {\bf 84}, 103508 (2011).

%\bibitem{boyce}
%W.~E.~Boyce and R.~C.~DiPrima, {\it Elementary differential equations and boundary value problems}, Wiley New York, (2004).

\bibitem{maartens}
R.~Maartens, {\it Causal Thermodynamics in Relativity}, arXiv:astro-ph/9702070.

\bibitem{prigogine}
I.~Prigogine, J.~Geheniau, E.~Gunzig and P.~Nardone, Gen.\ Rel.\ Grav. {\bf 21}, 767 (1989).

\bibitem{zim}
W.~Zimdahl, Month.\ Not.\ R.\ Astron.\ Soc. {\bf 288}, 665 (1997).

\bibitem{zimdahl}
W.~Zimdahl, J.~Triginer and D.~Pav\'on, Phys.\ Rev.\ D {\bf 54}, 6101 (1996).

\bibitem{lima}
J.~A.~S.~Lima, M.~O.~Calvao and I.~Waga, {\it Cosmology, Thermodynamics and Matter Creation}, arXiv:0708.3397 [astro-ph].

\bibitem{waga}
R.~O.~Ramos, M.~V.~dos~Santos and I.~Waga, Phys.\ Rev.\ D {\bf 89}, 083524 (2014).

\bibitem{stegun}
M.~Abramowitz and I.~A.~Stegun, {\it Handbook of Mathematical Functions with Formulas, Graphs, and Mathematical Tables}, Dover, (1972).
\end{thebibliography}
\end{document}